\documentclass[preprint,journal]{vgtc}       % preprint (journal style)

\ifpdf%                                % if we use pdflatex
\pdfoutput=1\relax                   % create PDFs from pdfLaTeX
\usepackage{graphicx}                % allow us to embed graphics files
\DeclareGraphicsExtensions{.pdf,.png,.jpg,.jpeg} % for pdflatex we expect .pdf, .png, or .jpg files
\else%                                 % else we use pure latex
\usepackage{graphicx}                % allow us to embed graphics files
\DeclareGraphicsExtensions{.eps}     % for pure latex we expect eps files
\fi%

\graphicspath{{figures/}{pictures/}{images/}{./}} % where to search for the images

\usepackage{microtype}                 % use micro-typography (slightly more compact, better to read)
\PassOptionsToPackage{warn}{textcomp}  % to address font issues with \textrightarrow
\usepackage{textcomp}                  % use better special symbols
\usepackage{mathptmx}                  % use matching math font
\usepackage{times}                     % we use Times as the main font
\usepackage{cite}                      % needed to automatically sort the references
\usepackage{tabu}                      % only used for the table example
\usepackage{booktabs}                  % only used for the table example
\usepackage{wrapfig}
\usepackage{todonotes}
\usepackage{mathptmx}
\usepackage{graphicx}
\usepackage{times}
\usepackage{amsmath,amscd,amssymb, amsthm}
\usepackage{times}
\usepackage[abs]{overpic}
\usepackage{cases}
\usepackage{float}
\usepackage{caption}
\usepackage{subfloat}
\usepackage{subfig}
\usepackage{wrapfig}
\usepackage{placeins}
\usepackage{multirow}
\usepackage{paralist}

\onlineid{1426}

\long\def\symbolfootnote[#1]#2{\begingroup%
\def\thefootnote{\fnsymbol{footnote}}\footnote[#1]{#2}\endgroup}

\DeclareMathOperator{\trace}{trace}
\DeclareMathOperator{\minor}{minor}
\DeclareMathOperator{\arcosh}{arcosh}
\DeclareMathOperator{\arsinh}{arsinh}
\DeclareMathOperator{\sign}{sign}

\vgtccategory{Research}
\vgtcpapertype{algorithm/technique}

\title{Feature Curves and Surfaces of 3D Asymmetric Tensor Fields}

\author{
Shih-Hsuan Hung,
Yue Zhang, \textit{Member,~IEEE},
Harry Yeh,
Eugene Zhang, \textit{Senior Member,~IEEE}
}

\authorfooter{
	\item
	S. Hung is with the School of Electrical Engineering and Computer Science, Oregon State University. E-mail: hungsh@oregonstate.edu.
	\item
	Y. Zhang is an Associate Professor with the School of Electrical Engineering and Computer Science, Oregon State University. E-mail: zhangyue@oregonstate.edu.
	\item
	H. Yeh is a Professor with the School of Civil and Construction Engineering, Oregon State University. E-mail: harry@engr.orst.edu.
	\item
	E. Zhang is a Professor with the School of Electrical Engineering and Computer Science, Oregon State University. E-mail: zhange@eecs.oregonstate.edu.
	
}

\shortauthortitle{Hung \MakeLowercase{\textit{et al.}}: Feature Curves and Surfaces of 3D Asymmetric Tensor Fields}
\abstract{
	3D asymmetric tensor fields have found many applications in science and engineering domains, such as fluid dynamics and solid mechanics. 3D asymmetric tensors can have complex eigenvalues, which makes their analysis and visualization more challenging than 3D symmetric tensors. Existing research in tensor field visualization focuses on 2D asymmetric tensor fields and 3D symmetric tensor fields. In this paper, we address the analysis and visualization of 3D asymmetric tensor fields. We introduce six topological surfaces and one topological curve, which lead to an eigenvalue space based on the tensor mode that we define. In addition, we identify several non-topological feature surfaces that are nonetheless physically important. Included in our analysis are the realizations that triple degenerate tensors are structurally stable and form curves, unlike the case for 3D symmetric tensors fields. Furthermore, there are two different ways of measuring the relative strengths of rotation and angular deformation in the tensor fields, unlike the case for 2D asymmetric tensor fields. We extract these feature surfaces using the A-patches algorithm. However, since three of our feature surfaces are quadratic, we develop a method to extract quadratic surfaces at any given accuracy. To facilitate the analysis of eigenvector fields, we visualize a hyperstreamline as a tree stem with the other two eigenvectors represented as thorns in the real domain or the dual-eigenvectors as leaves in the complex domain. To demonstrate the effectiveness of our analysis and visualization, we apply our approach to datasets from solid mechanics and fluid dynamics.
} % end of abstract


\CCScatlist{ % not used in journal version
	\CCScat{K.6.1}{Management of Computing and Information Systems}%
	{Project and People Management}{Life Cycle};
	\CCScat{K.7.m}{The Computing Profession}{Miscellaneous}{Ethics}
}

\teaser{
	\centering
	\begin{overpic}[width={\textwidth}]{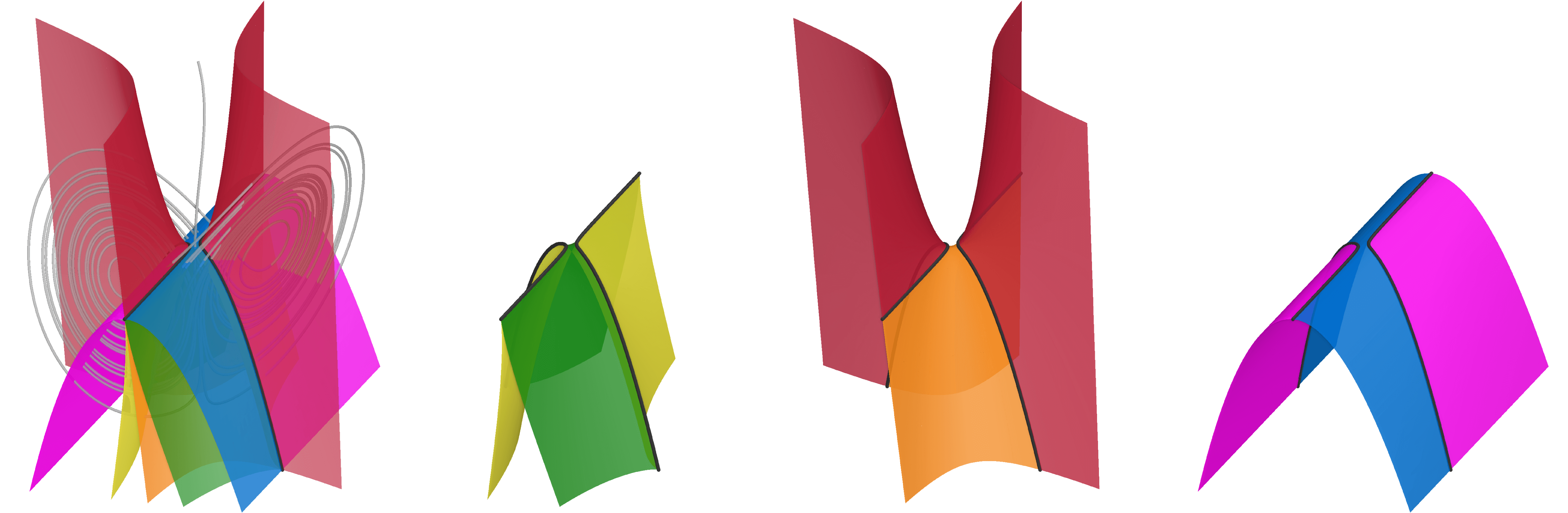}
		\put(28, -7)   {\small (a) all features}
		\put(130, -7)  {\small (b) degenerate surfaces}
		\put(260, -7)  {\small (c) neutral surfaces}
		\put(370, -7)  {\small (d) balanced surfaces}
	\end{overpic}
	\caption{
		We introduce a number of topological feature curves and surfaces for 3D asymmetric tensor fields, such as the velocity gradient tensor of the Lorenz attractor~\cite{lorenz:1963:deterministic} (a).
		These surfaces include (b) linear degenerate surfaces (green) and planar degenerate surfaces (yellow), (c) real neutral surfaces (orange) and complex neutral surfaces (red), and (d) linear balanced surfaces (blue) and planar balanced surfaces (magenta).
		Note that all of these surfaces intersect exactly at the triple degenerate curves (black). Furthermore, these surfaces exhibit a two-way rotational symmetry.
		Our analysis leads to an eigenvalue space for the analysis of 3D asymmetric tensor fields (Section~\ref{sec:eigenvalue_manifold}).
		In addition, our topological feature surfaces separate the two critical points in the attractor where steady convection occurs ((a): the centers of the swirls in the butterfly-shaped trajectory).
	}
	\label{fig:teaser}
}

\vgtcinsertpkg

\begin{document}

	%% The ``\maketitle'' command must be the first command after the
	%% ``\begin{document}'' command. It prepares and prints the title block.
	
	%% the only exception to this rule is the \firstsection command
	\firstsection{Introduction}\label{sec:introduction}
	
	\maketitle
	
	%% \section{Introduction} %for journal use above \firstsection{..} instead
	
\begin{figure}[!t]
		\centering
		\begin{overpic}[width={0.95\columnwidth}]{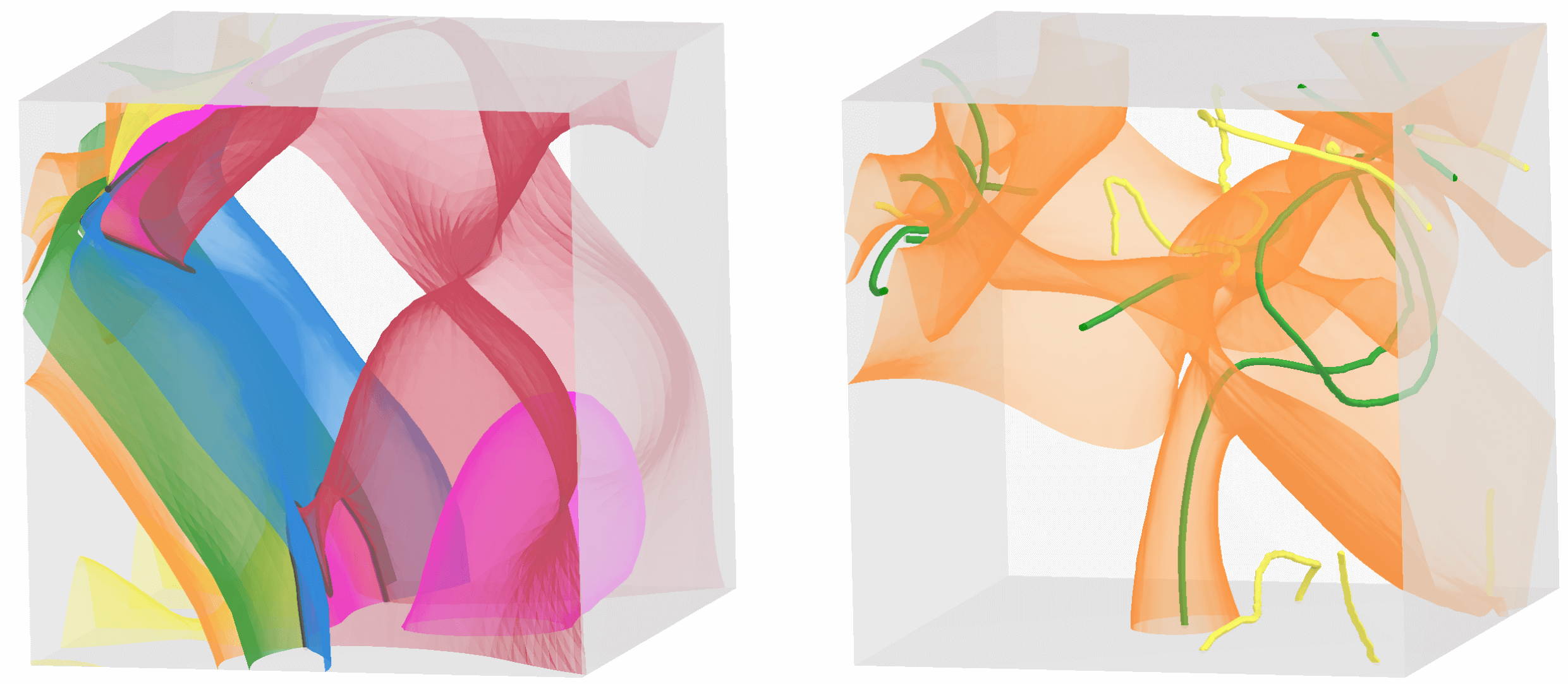}
%			\put(5, -7) {(a) asymmetric tensor field}
%			\put(125, -7) {(b) symmetric part}
		\end{overpic}
		\caption{%{\bf Topological features in 3D asymmetric tensor fields and 3D symmetric tensor fields.}
			The rich structure in a 3D asymmetric tensor field, such as that of the velocity gradient tensor of the Rayleigh-B{\'e}rnard flow (left) cannot be adequately captured when only visualizing the symmetric part of the tensor field (right).
		}
		\label{fig:comp_Sym}
	\end{figure}

	%
	Asymmetric tensor fields appear in many scientific and engineering applications. In fluid dynamics, the gradient of the velocity is an asymmetric tensor field that encodes fundamental behaviors such as rotation, angular deformation (also known as pure shear), and volumetric deformation. Similar behaviors are encoded in the deformation gradient tensor in solid mechanics. While these types of motions can be understood by visualizing the vector field itself, tensor field visualization provides a more direct visual representation~\cite{Chen:11}.
	Existing visualization techniques for 3D asymmetric tensor fields have focused on three different approaches.
	First, the tensor field is analyzed locally, with a focus on designing proper glyph representations for tensors~\cite{gerrits:2016:glyphs}.
	Second, the topological structures of the symmetric part of the tensor field are extracted and visualized~\cite{Palacios:16}.
	However, asymmetric tensors can have complex eigenvalues which lead to features and structures that are not well preserved by the symmetric part of the tensor field (Figure~\ref{fig:comp_Sym}).
	Third, researchers have attempted to understand 3D asymmetric tensor fields by performing 2D analysis on the projection of the tensor field on some probe planes or surfaces.
	Unfortunately, where the projected tensors have complex eigenvalues does not usually coincide with the complex domain of the original 3D tensor field.
	These difficulties highlight a fundamental need to perform topological analysis and visualization {\em directly} on 3D asymmetric tensor fields, rather than their projection on 2D or their symmetric part.

	In this paper, we introduce the notion of {\em tensor mode} for 3D asymmetric tensors, which leads to a model that we refer to as the {\em eigenvalue space}.
	Each point in the eigenvalue space gives rise to a levelset surface that we call a {\em mode surface}.
	We have identified seven special modes, which give rise to six topological feature surfaces and one feature curve: {\em linear and planar degenerate surfaces} (Figure~\ref{fig:teaser} (b)), {\em real and complex neutral surfaces} (Figure~\ref{fig:teaser} (c)), {\em linear and planar balanced surfaces} (Figure~\ref{fig:teaser} (d)), and {\em triple degenerate curves} (the black curve in (Figure~\ref{fig:teaser} (b-d)).
	In particular, unlike 3D {\em symmetric} tensor fields, triple degenerate points are stable features in 3D {\em asymmetric} tensor fields and form curves (triple degenerate curve).
	We also observe that, unlike 2D asymmetric tensor fields, there are two different measures for the relative strengths of rotation and shear in the tensor, emphasizing the significance of the degenerate surface and the balanced surface.
	These differences highlight the richer structures in 3D asymmetric tensor fields.
	
	In addition, we define some non-topological feature surfaces such as the levelsets of the tensor magnitude ({\em magnitude surfaces}) and isotropicity ({\em isotropicity surfaces}).
	
	To better understand the eigenvector behaviors in the asymmetric field, we develop an {\em augmented hyperstreamline} visualization method. When traveling along a hyperstreamline following one eigenvector field, we also visualize the other eigenvectors in the real domain and the dual-eigenvectors in the complex domain along the hyperstreamline. The hyperstreamline is shown as a tree stem, while the other eigenvectors in the real domain are visualized as thorns attached to the stem. Similarly, the dual-eigenvectors in the complex domain are visualized as leaves attached to the stem.
	This can be particularly useful for inspecting the eigenvector behavior when crossing special mode surfaces such as the neutral surface and the degenerate surface.
	For piecewise linear tensor fields defined on tetrahedral meshes, three of the aforementioned feature surfaces, such as the balanced surfaces, magnitude surfaces, and isotropicity surfaces are quadratic inside each tetrahedron.
	For such surfaces, we provide a quadratic surface extraction method that leads to a seamless extracted surface.
	For features surfaces of a higher-degree such as the neutral surface, the degenerate surface, and other mode surfaces, we employ the A-patches method~\cite{luk:2009}.
	Finally, we extract the triple degenerate curve by finding the intersection of the balanced surface and the neutral surface.
	
	We demonstrate the utility of our approach by applying our tensor field analysis and visualization to solid mechanics and fluid dynamics applications and providing physical interpretation.
	\section{Related Work}\label{sec:related_work}
	Tensor field visualization has advanced much in the last decades~\cite{EPFL-BOOK-138668,Kratz:13}. Topological analysis of tensor fields has found many applications in understanding solid and fluid mechanics data.
	Existing topology-driven tensor field visualization has focused on symmetric tensors of two- and three-dimensions. Tensor field topology is first studied by Delmarcelle and Hesselink\cite{delmarcelle:visualizing}, who extend the notions of singularities and separatrices from vector fields to 2D symmetric tensor fields.
	The topological features of 3D symmetric tensor fields are first studied by Hesselink et al.~\cite{hesselink:topology}, who define degenerate points as those where the tensor field has an eigenvalue with a multiplicity of {\em three}, i.e. triple degeneracy. Zheng and Pang~\cite{Zheng:04} point out that triple degenerate points are structurally unstable. That is, under an arbitrarily small perturbation to the tensor field such points disappear. Instead, Zheng and Pang define the topology of a 3D symmetric tensor field as the collection of {\em double degenerate points}, where the tensor field has two eigenvalues, one of which is repeating (multiplicity of two). Such points form curves, i.e. degenerate curves. Since then, a number of techniques have been developed to extract degenerate curves~\cite{Zheng:05a,Tricoche:08,Palacios:16,Roy:18}.
	In particular, Tricoche et al.~\cite{Tricoche:08} point out that the degenerate curves are a subset of the ridge and valley lines of {\em tensor mode}, a tensor invariant whose name originated from mechanics~\cite{CRISCIONE:00}. With this formulation, Tricoche et al.~\cite{Tricoche:08} introduce the concept of tensor mode to the Visualization community and the idea of using tensor mode to define and extract topological structures.
	More recently, a number of feature surfaces have been introduced for 3D symmetric tensor fields, such as neutral surfaces and mode surfaces~\cite{Palacios:16}, extremal surfaces~\cite{Zobel2017}, and fiber surfaces\cite{raith2019tensor}.

	The visualization of asymmetric tensor fields starts more recently, and it has focused on 2D. Zheng and Pang~\cite{Zheng:05c} extend the topological analysis from 2D symmetric tensor fields to 2D {\em asymmetric} tensor fields with the introduction of {\em dual-eigenvectors} in the {\em complex domains} where the tensor field has complex eigenvalues. Zhang et al.~\cite{Zhang:09} provide rigorous analysis of 2D asymmetric tensor fields with the introduction of the notion of {\em eigenvalue manifold}. Chen et al.~\cite{Chen:11} introduce a visualization in which glyphs and hyperstreamlines are both used in visualizing asymmetric tensor fields. Lin et al.~\cite{Lin:12} introduce the notions of {\em eigenvalue graphs} and {\em eigenvector graphs} for 2D asymmetric tensor fields, which are extended to surfaces and a multi-scale framework by Khan et al.~\cite{Khan:20}.
	Despite the advances in 3D symmetric tensor fields and 2D asymmetric tensor fields, there has been relatively little work in the topological analysis of 3D asymmetric tensor fields. Visualization research on such fields is usually focused on glyph design~\cite{gerrits:2016:glyphs}. In this paper, we provide the results of our initial investigation of the topological analysis for 3D asymmetric tensor fields.
	\begin{figure*}[!t]
		\centering
		\centering%
		\begin{overpic}[width={\textwidth}]{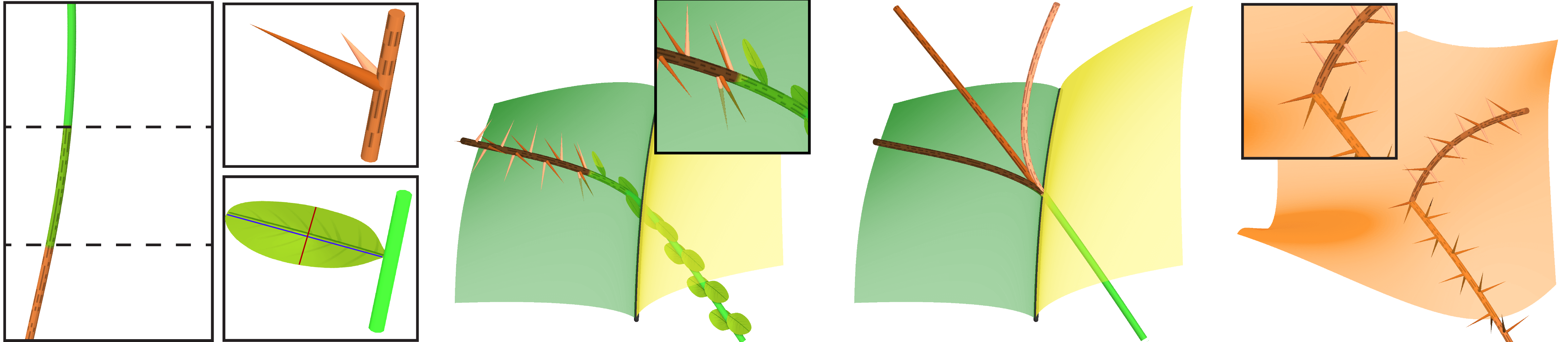}
			\put(65, -8)  {(a)}
			\put(30, 20)  {real}
			\put(30, 10)  {domain}
			\put(30, 60)  {outer}
			\put(30, 50)  {complex}
			\put(30, 40)  {domain}
			\put(30, 95)  {inner}
			\put(30, 85)  {complex}
			\put(30, 75)  {domain}
			\put(78, 30)  {$w_1$}
			\put(92, 15)  {$w_2$}
			\put(200, -8) {(b)}
			\put(330, -8) {(c)}
			\put(460, -8) {(d)}
		\end{overpic}
		\caption{
			We visualize a hyperstreamline following an eigenvector field as a tree stem, with the other eigenvectors in the real domain as thorns, and the dual-eigenvectors in the complex domain as leaves (a). A segment of the hyperstreamline is textured with a wood texture inside the real domain and given a smooth appearance in the inner complex domain. The segment inside the outer complex domain is a composition of the two appearances.
			When crossing the degenerate surface (b), the dominant eigenvector field (tree stem in this case) in the real domain changes into the real eigenvector field in the complex domain.
			Notice that the repeating eigenvector at the crossing point is the limit of the other two eigenvectors (thorns) from the real domain and the major dual-eigenvector (long axes of the leaves) from the complex domain.
			When crossing the triple degenerate curve (c), all three eigenvectors from the real domain converge to the only eigenvector at the triple degenerate point.
			Additionally, the dominant eigenvector field is discontinuous at the real neutral surface (d).
			%
			%We detail the hyperstreamlines in \sref{sec:hyperstreamlines}.
		}
		\label{fig:degenerate_crossing}
	\end{figure*}

	\section{Tensor Background}
	\label{sec:math_background}
	Before presenting our analysis, we first review relevant mathematical background on 3D asymmetric tensor fields.
	We start with 3D asymmetric tensors, which, under a given basis, can be represented as $3\times 3$ matrices.
	A $3 \times 3$ tensor has a {\em characteristic polynomial} $f(\lambda)=\lambda^3+a_2\lambda^2 + a_1\lambda+a_0$ such that $f(T)=0$. The {\em trace} of $T$ is $\trace(T)=-a_{2}$. When the trace is zero, the tensor $T$ is referred to as being {\em traceless}.
	The {\em determinant} of $T$ is $\det(T)= -a_0$, and the {\em minor} is $\minor(T)=a_1$.
	Additionally, the set of all $3 \times 3$ tensors form a $9$-dimensional linear space, on which the following inner product of two tensors $R$ and $S$ can be introduced~\cite{spence2000elementary}: $\langle R, S \rangle = \sum_{i=1}^{3}\sum_{j=1}^{3}R_{ij}S_{ij} = \trace(S^T R)$.
	With this product, one can define the {\em magnitude} of a tensor $T$ as $||T||=\sqrt{\langle T, T \rangle}$.
	The roots of the characteristic polynomial $f(\lambda)$ are the eigenvalues of $T$. There are either three mutually distinct real-valued eigenvalues, one real-valued eigenvalue and two complex-valued conjugate eigenvalues, two real-valued eigenvalues with one of them having a multiplicity of two, or one real-valued eigenvalue of multiplicity of three.
	When all three eigenvalues are real and mutually distinct, we refer to their largest, middle, and smallest eigenvalues as the {\em major}, {\em medium}, and {\em minor eigenvalues}, respectively.
	When there is only one real eigenvalue, it is referred to as the {\em real eigenvalue} of $T$.
	When there are two eigenvalues, we refer to the eigenvalue of a multiplicity of two as the {\em repeating eigenvalue} and the other eigenvalue the {\em dominant eigenvalue}. The notions of {\em major eigenvectors}, {\em medium eigenvectors}, {\em minor eigenvectors}, {\em real eigenvectors}, {\em repeating eigenvectors}, and {\em dominant eigenvectors} can be defined as the eigenvectors corresponding to the respective eigenvalues.
	A tensor $T$ is {\em symmetric} if it is equal to its transpose; otherwise, it is {\em asymmetric}. A special case of asymmetric tensors is {\em anti-symmetric tensors}, which are equal to their negated transpose. The eigenvalues of a symmetric tensor are guaranteed to be real-valued, while the eigenvalues of an asymmetric tensor can be either real-valued or complex-valued. 	
	Furthermore, eigenvectors belonging to different eigenvalues form an orthonormal basis for symmetric tensors.
	For asymmetric tensors, even when the eigenvalues are real-valued, their respective eigenvectors are not mutually perpendicular.

	A tensor field is a continuous tensor-valued function defined in the domain. A hyperstreamline is a curve that is tangent to an eigenvector field everywhere along its path. For example, a {\em dominant hyperstreamline} follows the dominant eigenvector field, while a {\em real hyperstreamline} follows the real eigenvector field.
	%%%%%%%%%%%%%%%%%%%%%%%%%%%%%%%%%%%%%%%%%%%%%%%%%%%%%%%%%%%%%%%%%%%%%%%%%%%%%%%%%%%%%%%%%%%%%%%%%%%%%%%%%%%%%%%%%%

	\section{Analysis of 3D Asymmetric Tensor Fields }
	\label{sec:eigenvalue_manifold}
	In this section, we describe our analysis of 3D asymmetric tensor fields.
	An $3 \times 3$ asymmetric tensor $T$ can be uniquely decomposed as
	\begin{equation}
		T = D+A,
		\label{eq:decomposition}
	\end{equation}
	where $D=\frac{\trace(T)}{3}\mathbb{I}$ is a multiple of the identity matrix $\mathbb{I}$ and $A=T-D$ is a traceless tensor that is referred to the {\em deviator} of $T$.
	Note that $T$ and $A$ have the same set of eigenvectors, i.e. anisotropy.
	Therefore, we begin with the analysis of 3D traceless asymmetric tensors in the following subsections.

	\subsection{Dual-Eigenvectors}
	A traceless asymmetric tensor with complex eigenvalues has a real Schur form~\cite{Aldhaheri:89} of $\begin{pmatrix}
		a & -c & \quad\quad d \\
		c &  \quad b & \quad\quad e \\
		0 &  \quad 0 & -a-b
	\end{pmatrix}$
	%\begin{equation*}
	%\begin{pmatrix}
	%	a & -c & \quad\quad d \\
	%	c &  \quad b & \quad\quad e \\
	%	0 &  \quad 0 & -a-b
	%\end{pmatrix},
	%\end{equation*}
	%
	where $(a-b)^2<4c^2$ under some orthonormal basis $\langle v_1, v_2, v_3 \rangle$.
	$T$ has one real eigenvalue $\lambda_3=-a-b$ and two complex eigenvalues $\lambda_{1,2}=\frac{(a+b)\pm \sqrt{4c^2-(a-b)^2}i}{2}$.
	Note that the eigenvectors corresponding to the complex eigenvalues are also complex-valued.
	We extend the notion of {\em dual-eigenvectors} from 2D asymmetric tensor fields~\cite{Zheng:05c} to 3D.
	In the plane spanned by $\langle v_1, v_2 \rangle$, the projection of $T$ has the form $\begin{pmatrix} a & -c  \\ c &  \quad b \end{pmatrix}$, whose major and minor dual-eigenvectors are well-defined.
	We refer to the dual eigenvectors of the projection tensor as the dual-eigenvectors of $T$.

	\subsection{Degenerate Surface}
	Given a 3D asymmetric tensor field, the set of points in the tensor field with three mutually distinct real eigenvalues is referred to as the {\em real domain} of the field, while the set of points with one real eigenvalue and two complex conjugate eigenvalues is referred to as the {\em complex domain} of the field. The boundary between the real domain and the complex domain consists of points with one real-valued eigenvalue with a multiplicity of at least two. We refer to such a boundary point as a {\em degenerate point}.
	The real Schur form for degenerate traceless tensors is expressed as $\begin{pmatrix}
		a & c & \quad d \\
		0 & a & \quad e \\
		0 & 0 & -2a
	\end{pmatrix}.$
	%\begin{equation*}
	%\begin{pmatrix}
	%	a & c & \quad d \\
	%	0 & a & \quad e \\
	%	0 & 0 & -2a
	%\end{pmatrix}.
	%\end{equation*}
	%
	When $a=0$, $T$ has one real eigenvalue $0$ with a multiplicity of three.
	It is therefore referred to as a {\em triple degenerate tensor}.
	Otherwise, $T$ has one real eigenvalue $a$ with a multiplicity of two and another real eigenvalue $-2a$. In this case, $T$ is a {\em double degenerate tensor}.
	Furthermore, for a double degenerate tensor, the $2\times 2$ sub-block corresponds to a plane, and the projection of the tensor onto the plane is a 2D degenerate tensor.
	In general, a traceless tensor is degenerate if and only if its {\em discriminant} $\Delta(T)=0$ where $\Delta(T)=-27\det(T)^2-4\minor(T)^3$.
	Note that the discriminant $\Delta$ can be negative for asymmetric tensors.
	Consequently, the set of degenerate points is co-dimension one and forms a surface which we refer to as the {\em degenerate surface}.
	Additionally, the set of triple degenerate tensors has one additional constraint which is $a=0$. Therefore, this set of tensors forms curves, i.e. {\em triple degenerate curve}.
	Notice that in 3D symmetric tensor fields, the complex domain is empty, triple degenerate tensors are structurally unstable, and double degenerate tensors form curves.
	Contrasting these properties with the properties of 3D asymmetric tensor fields suggests that features in a 3D asymmetric tensor field cannot be properly represented by the features in its symmetric part~\cite{Palacios:16}.
	We wish to understand the eigenvector behavior at the degenerate surface.
	For this, we travel along a dominant hyperstreamline towards the degenerate surface as shown in Figure~\ref{fig:degenerate_crossing}.
	Since there are two more eigenvectors in the real domain and two dual-eigenvectors in the complex domain, we develop a visualization metaphor in which the hyperstreamline is the stem of a plant to which thorns and leaves can be attached.
	Along the stem, the other eigenvectors are represented as thorns and the dual-eigenvectors as leaves (Figure~\ref{fig:degenerate_crossing} (a)). We refer to a hyperstreamline with thorns and leaves as an {\em augmented hyperstreamline}.
	Notice that when traveling along the dominant hyperstreamline towards the degenerate surface from the real domain (Figure~\ref{fig:degenerate_crossing} (b)), the other two eigenvectors converge and become the same at the degenerate surface, which is the repeating eigenvector.
	On the other hand, when traveling from the complex domain towards the degenerate surface, the eccentricities of the leaves increase towards one (the ellipse becomes a thin line).
	The major dual-eigenvectors converge to the repeating eigenvector at the degenerate surface.
	It is also possible to cross the triple degenerate curve.
	In this case (Figure~\ref{fig:degenerate_crossing} (c)), all three eigenvectors in the real domain converge to the only eigenvector at the triple degenerate curve (the three stems become tangent at their common intersection point).
	On the other side, the real eigenvector from the complex domain also converges to the same eigenvector at the triple degenerate curve.
	These behaviors at the degenerate surface and triple degenerate curve signify their topological importance.
	\subsection{Neutral Surface}
	In the real domain, a 3D asymmetric traceless tensor $T$ has three mutually distinct real eigenvalues (i.e. $\lambda_1>\lambda_2>\lambda_3$) which sum to zero. There are three cases:
	\begin{inparaenum}[i)]
		\item {\em linear}: $\lambda_2<0$ where the major eigenvalue $\lambda_1$ is the {\em dominant eigenvalue},
		\item {\em planar}: $\lambda_2>0$ where the minor eigenvalue $\lambda_3$ is the dominant eigenvalue, and
		\item {\em neutral}: $\lambda_2=0$ where the dominant eigenvalue is not well-defined, since the major eigenvalue and minor eigenvalue have an equal absolute value but opposite signs.
	\end{inparaenum}
	
	Similarly, we can classify degenerate traceless tensors as being {\em linear}, {\em planar}, or {\em neutral} if the repeating eigenvalue (corresponding to $\lambda_2$ in the real domain) is negative, positive, or zero, respectively.
	Note that the set of neutral degenerate tensors is exactly the set of triple degenerate tensors. Furthermore, the dominant eigenvalue of a degenerate tensor is positive (linear), negative (planar), and not well-defined (neutral).
	In the complex domain, we also classify tensors in a similar fashion.
	Such tensors have only one real eigenvalue, which is the dominant eigenvalue. We refer to such a tensor as linear, planar, or neutral if the real eigenvalue is positive, negative, or zero, respectively.
	Note that this classification of linearity/planarity/neutrality is consistent with that for the real domain and degenerate surface.
	That is, when travelling along a path from the real domain to the complex domain without ever reaching any neutral tensor, the linearity/planarity does not change.
	The real Schur form of real neutral tensors is $\begin{pmatrix}
		a & \quad c  & d \\
		0 & -a & e \\
		0 & \quad 0  & 0
	\end{pmatrix}.$
	%\begin{equation*}
	%\begin{pmatrix}
	%	a & \quad c  & d \\
	%	0 & -a & e \\
	%	0 & \quad 0  & 0
	%\end{pmatrix}.
	%\end{equation*}
	%
	The projection of the tensor onto the plane spanned by the major and minor eigenvectors is traceless and has two real eigenvalues $\pm a$.
	On the other hand, the real Schur form of complex neutral tensors is $\begin{pmatrix}
		a & -c & d \\
		c & -a & e \\
		0 & \quad 0  & 0
	\end{pmatrix}.$
	%\begin{equation*}
	%\begin{pmatrix}
	%	a & -c & d \\
	%	c & -a & e \\
	%	0 & \quad 0  & 0
	%\end{pmatrix}.
	%\end{equation*}
	%
	The projection of such a tensor onto the plane spanned by the dual-eigenvectors is also traceless and has a pair of conjugate complex eigenvalues.
	The collection of real neutral tensors, triple degenerate tensors, and complex neutral tensors form a surface which we refer to as the {\em neutral surface}. It separates the domain of the tensor field into the {\em linear domain} and the {\em planar domain}. Furthermore, the neutral surface is characterized by $\det(T)=0$.
	When one travels from the linear domain into the planar domain through the real neutral surface, the dominant eigenvalue (and eigenvector) switches from the major eigenvalue (and eigenvector) to the minor eigenvalue (and eigenvector). Notice the sudden change in the hyperstreamline direction in Figure~\ref{fig:degenerate_crossing} (d).
	Furthermore, the degenerate surface intersects the neutral surface at exactly the triple degenerate curve.

	\subsection{Balanced Surface}
	A traceless asymmetric tensor $T$ can be uniquely decomposed as the sum of a symmetric tensor $S$ and an anti-symmetric tensor $R$. When $T$ is the velocity gradient of an incompressible flow, $S$ represents the rate of angular deformation and $R$ the rate of rotation in the fluids.
	We define the strength of rotation as $\tau_R=\sqrt{\langle R, R \rangle}$ and the strength of the angular deformation (shear) as $\tau_S=\sqrt{\langle S, S \rangle}$, respectively.
	A tensor $T$ is {\em shear-dominant} if $\tau_S>\tau_R$.
	On the other hand, $T$ is {\em rotation-dominant} if $\tau_S<\tau_R$. When $\tau_S=\tau_R$, we refer to $T$ as a {\em balanced tensor}.
	In the 2D case, a $2\times 2$ tensor $T$ has complex eigenvalues if and only if $\tau_R>\tau_S$.
	Otherwise, it has only real eigenvalues.
	That is, the complex domain is identical to the rotation-dominant domain, and the real domain is identical to the shear-dominant domain.
	However, the situation is different in 3D. It can be verified that a tensor $T$ is balanced, i.e. $\tau_R=\tau_S$, if and only if $\minor(T)=0$.
	The real Schur form for a balanced tensor $T$ is $\begin{pmatrix}
		a & -c & \quad\quad d \\
		c & \quad b & \quad\quad e \\
		0 & \quad 0 & -a-b
	\end{pmatrix}$
	%\begin{equation*}
	%\begin{pmatrix}
	%	a & -c & d \\
	%	c & b & e \\
	%	0 & 0 & -a-b
	%\end{pmatrix},
	%\end{equation*}
	where $c^2=a^2+ab+b^2$.
	In this case, $|a|$, $|b|$, and $|c|$ form the side lengths of a triangle with the angle between sides of $|a|$ and $|b|$ being $120^\circ$ if $ab>0$ and $60^\circ$ if $ab<0$. Furthermore, $T$ is linear if $a+b<0$ and planar if $a+b>0$.
	Note that a balanced tensor $T$ must have complex eigevalues except when $a=b=0$, i.e. the tensor is triple degenerate.
	Consequently, the set of balanced tensors is not the same as the set of degenerate tensors. That is, the complex domain is not the same as the rotation-dominant domain for 3D asymmetric tensor fields.
	The balanced surface divides the complex domain into
	\begin{inparaenum}[i)]
		\item {\em inner complex domain} (dominated by rotation), and
		\item {\em outer complex domain} (dominated by shear).
	\end{inparaenum}
	This signifies the importance of the balanced surface as a feature in the tensor field.
	Moreover, the difference between the balanced surface and the degenerate surface shows the richer structure in 3D asymmetric tensor fields when it comes to understanding the interaction between rotation and shearing.
	Another important observation is that the neutral surface, the degenerate surface, and the balanced surface intersect exactly at the triple degenerate curve, signifying the latter's topological importance.

	\subsection{Tensor Mode}
	An important invariant for 3D symmetric tensor fields is their mode~\cite{Palacios:16}, which is intricately connected to the topology of the fields.
	For example, the mode is zero at precisely the neutral surface and $\pm 1$ at precisely the degenerate curves.
	All possible mode values for 3D symmetric tensors are between $-1$ and $1$.

	Extending the notion of tensor modes from the symmetric case, we define the mode for 3D asymmetric tensors in a way that is motivated by the formulas for the eigenvalues of the tensor.
	Let $T$ be a traceless tensor.
	When the discriminant $\Delta(T) \ge 0$, $T$ has three real eigenvalues:
	\begin{equation} \label{eq:real_domain}
		\begin{split}
			\lambda_1 &= -2\sqrt{-\frac{p}{3}} \cos \left(\frac{1}{3}\arccos\left(\frac{3q}{2p}\sqrt{\frac{-3}{p}}\right)+\frac{2\pi}{3}\right) \\
			\lambda_2 &= -2\sqrt{-\frac{p}{3}} \cos \left(\frac{1}{3}\arccos\left(\frac{3q}{2p}\sqrt{\frac{-3}{p}}\right)\right) \\
			\lambda_3 &= -2\sqrt{-\frac{p}{3}} \cos \left(\frac{1}{3}\arccos\left(\frac{3q}{2p}\sqrt{\frac{-3}{p}}\right)-\frac{2\pi}{3}\right), \\
		\end{split}
	\end{equation}
	where $p=\minor(T)$ and $q=\det(T)$. In particular, when $\Delta(T)=0$, two of the eigenvalues are the same and $T$ is degenerate.
	When $\Delta(T)<0$, $T$ is in the complex domain and the real eigenvalue can be expressed as
	\begin{equation}
		\lambda_1 = -\left\{
		\begin{array}{ll}
			2\frac{|q|}{q}\sqrt{-\frac{p}{3}}\cosh\left( \frac{1}{3} \arcosh\left(\frac{-3|q|}{2p}\sqrt{\frac{-3}{p}}  \right) \right) & \textrm{ if $p<0$} \\[1em]
			-\sqrt[3]{q} & \textrm{ if $p=0$} \\[0.5em]
			2\sqrt{\frac{p}{3}} \sinh\left(\frac{1}{3} \arsinh \left(\frac{3q}{2p}\sqrt{\frac{3}{p}}  \right)  \right) & \textrm{ if $p>0$}
		\end{array} \right.  \label{eq:complex_domain}
	\end{equation}
	From these formulas, we can see that the eigenvalues of $T$ are the result of the interplay among $p$, $q$, and $\Delta$. Therefore, we define the mode of $T$ as the triple $(\mu, \sign(p), \sign(q))$ where $\mu = \frac{3|q|}{2|p|}\sqrt{\frac{3}{|p|}}$.
	
Note that in the real domain, $\mu \in [0,1]$ and $p<0$.
	In the complex domain, $\mu\in (1,\infty)$ when $p<0$.
	Once $p > 0$, $T$ must be in the complex domain and $\mu \in [0,\infty)$.
	Given a particular mode $(\mu, \sign(p), \sign(q))$ for $\mu \in [0,\infty]$, the set of points in the tensor field of this mode forms a surface which we refer to as the {\em mode surface} of mode $(\mu, \sign(p), \sign(q))$.
	Note that feature surfaces that we have defined earlier have unique tensor modes.
	More specially,
	\begin{itemize}
		\item {\em Real neutral surface}: $(\mu=0, \sign(p)=\text{``-''}, \sign(q)=\text{``0''})$,
		\item {\em Complex neutral surface}: $(\mu=0, \sign(p)=\text{``+''}, \sign(q)=\text{``0''})$,
		\item {\em Linear degenerate surface}: $(\mu=1, \sign(p)=\text{``-''}, \sign(q)=\text{``+''})$,
		\item {\em Planar degenerate surface}: $(\mu=1, \sign(p)=\text{``-''}, \sign(q)=\text{``-''})$,
		\item {\em Linear balanced surface}: $(\mu=\infty, \sign(p)=\text{``0''}, \sign(q)=\text{``+''})$,
		\item {\em Planar balanced surface}: $(\mu=\infty, \sign(p)=\text{``0''}, \sign(q)=\text{``-''})$.
	\end{itemize}
	Another special mode is when $\sign(p)=\text{``0''}$ and $\sign(q)=\text{``0''}$; in this case, $\mu$ is undefined.
	The set of points with this mode is precisely the triple degenerate curve.
	\begin{figure}[!b]
		\centering%
		\includegraphics[width={\columnwidth}]{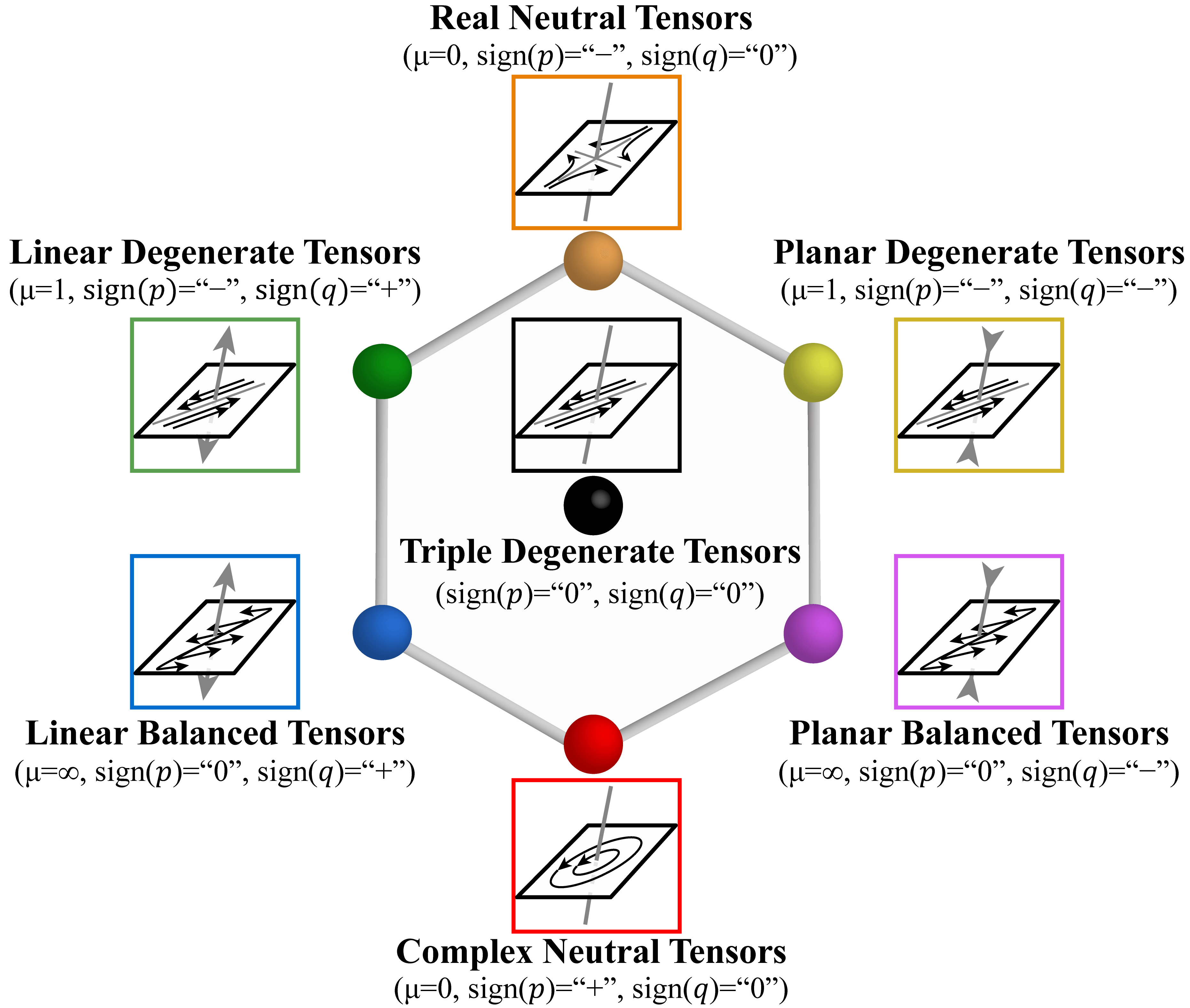}
		\caption{
			The eigenvalue space contains seven special tensors based on the tensor mode: real neutral tensors, complex neutral tensors, linear degenerate tensors, planar degenerate tensors, linear balanced tensors, planar balanced tensors, and triple degenerate tensors.
			When the tensor represents the velocity gradient of some incompressible flow, the corresponding flow patterns are illustrated next to the tensor.
			Notice that the flow pattern inside the 2D plane is simple shear when the tensor is degenerate (linear, planar, and triple).
			For neutral tensors, the projection is either a saddle (real neutral) or an elliptical pattern (complex neutral).
			For linear tensors, the flow leaves the plane in the third dimension, while for planar tensors, the flow enters the plane in the third dimension.
		}
		\label{fig:eigenvalue_manifold}
	\end{figure}
	\subsection{Eigenvalue Space}
	The definition of tensor mode allows us to construct a model for all 3D asymmetric tensors, which we refer to as the {\em eigenvalue space} for 3D asymmetric tensors.
	We first consider the set of all traceless tensors, which we map to the border of a hexagon and its center as shown in ~Figure~\ref{fig:eigenvalue_manifold}.
	Each point on the border of the hexagon represents a unique tensor mode.
	Starting from the top and continuing counterclockwise, we encounter special modes in the order of real neutral tensors, linear degenerate tensors, linear balanced tensors, complex neutral tensors, planar balanced tensors, and planar degenerate tensors, as shown in ~Figure~\ref{fig:eigenvalue_manifold}.
	In addition, the center of the hexagon corresponds to the triple degenerate tensors.
	On the other hand, points inside the hexagon other than the center do not correspond to any valid tensor mode.
	Note that the real domain consists of two edges in the hexagon (upper-left and upper-right), while the complex domain consists of the other four edges.
	The left and right edges correspond to the outer complex domain, while the {lower-left and lower-right edges correspond to the inner complex domain.
		Furthermore, the left half and the right half of the hexagon signify the symmetry between linear and planar tensors.

Note that the triple degenerate curve is adjacent to every mode surface. The domain of the tensor field is the disjoint union of all the mode surfaces. We can consider the volume being a book with each mode surface as a page and the triple degenerate curve as the book spine. Figure~\ref{fig:teaser} (a) illustrates this idea with a tensor field. 		

		For a tensor $T$ that may have a non-zero trace, we define the notion of {\em isotropicity} as
		\begin{equation}\label{eq:isotropicity}
			\eta(T)=\frac{\trace(T)}{\sqrt{3}||T||}
		\end{equation}
		Note that the isotropicity of a tensor must be between $-1$ and $1$, where the isotropicity of $\pm 1$ corresponds to a positive and negative multiple of the identity matrix, respectively.
		Given a 3D asymmetric tensor of unit magnitude, its mode and isotropocity uniquely determine its eigenvalues.
		In addition, when a tensor has an isotropicity of $\pm 1$, its deviator is zero, making its mode not well-defined.
		Consequently, we add to our eigenvalue space two additional special tensors: isotropicity of $1$ and isotropicity of $-1$.
		This leads to a double hexagonal cone and line segment in the middle as shown in ~Figure~\ref{fig:eigenvalue_manifold_cone}.
		The base of the double cone corresponds to traceless tensors, which are modeled by the hexagon in~Figure~\ref{fig:eigenvalue_manifold}.
		The top and bottom tip points in the double cone correspond to the $1$ and $-1$ isotropocities, respectively.
		Each point on the surface of this double cone as well as the line between the top and bottom tips corresponds to a unique combination of eigenvalues in a tensor up to a positive multiple.
		We refer to the double cone and the center line segment as the {\em eigenvalue space} for 3D asymmetric tensors.
		Note that pure isotropocity tensors (i.e. $\eta(T) = \pm 1$) are co-dimension eight in the space of 3D asymmetric tensors.
		Thus, they are structurally unstable.
		However, we include them in our eigenvalue space for their theoretical values.
		The set of points in the field with a given isotropicity $\eta\ne \pm 1$ is a surface, which we refer to as an {\em isotropicity surface}. A special isotropicity surface is the {\em traceless surface}, whose corresponding isotropicity value is zero. Note that for incompressible fluid data, its gradient tensor is always {\em traceless}; thus, the traceless surface becomes the whole domain.
		Another feature surface that we visualize is the {\em magnitude surface}, which consists of the points in the field with the same tensor magnitude that is not zero.
		Note that the set of zero magnitude tensors is co-dimension nine in the tensor field and thus structurally unstable.
		\begin{figure}[!t]
			\centering%
			\begin{overpic}[width={0.8\columnwidth}]{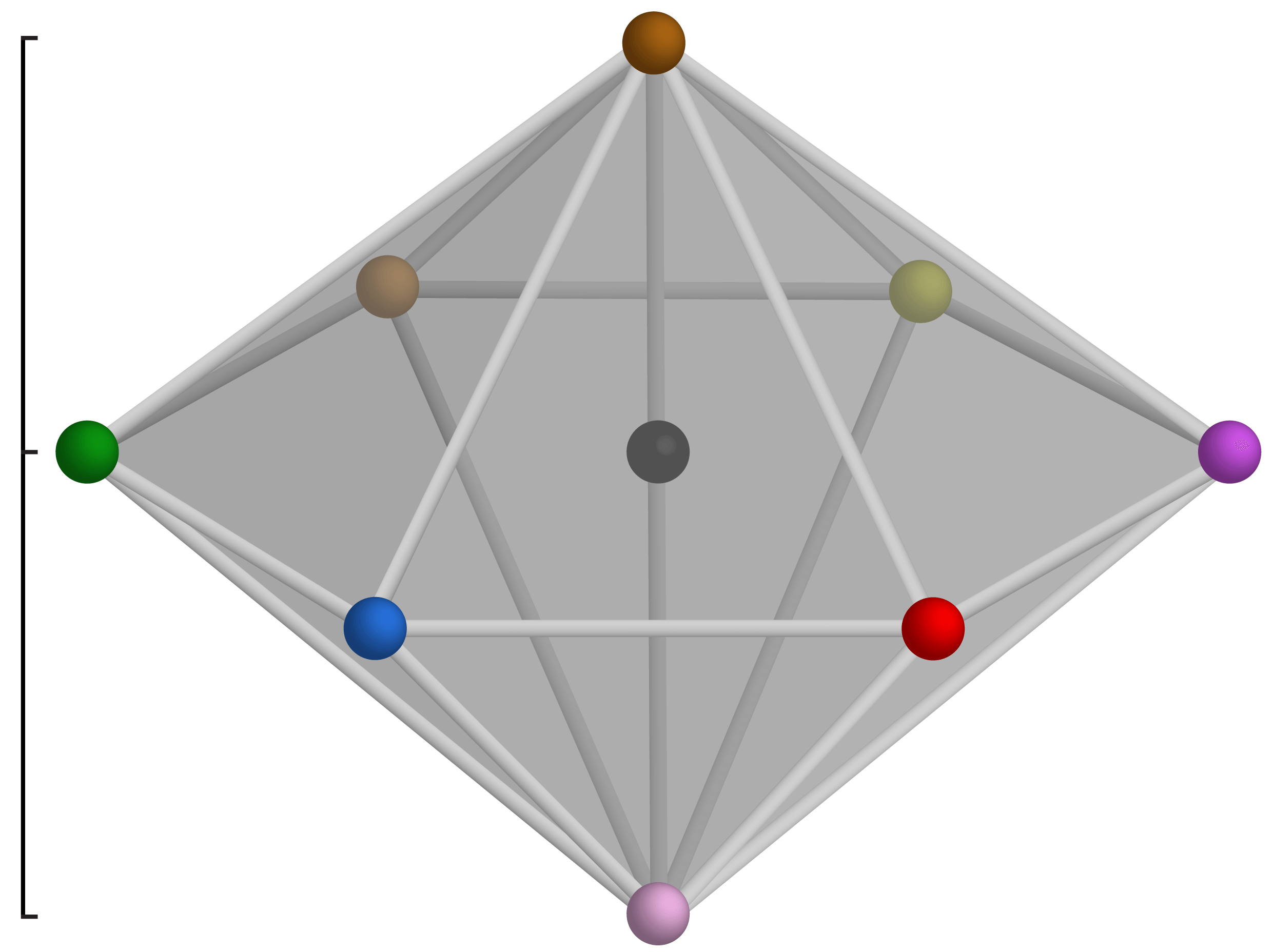}
				\put(-9,  5)  {-1}
				\put(-8, 55)  {0}
				\put(-8,100)  {1}
				\put(-12, -5)  {\small{Isotropicity}}
			\end{overpic}
			\caption{
				We add to our eigenvalue space two additional special tensors with isotropicity of $\pm 1$ at the top (brown dot) and the bottom (pink dot) and model the eigenvalue space as a hexagonal double cone with one additional line segment.}
			\label{fig:eigenvalue_manifold_cone}	
		\end{figure}
		\begin{figure*}[!t]
			\centering%
			\begin{overpic}[width={\textwidth}]{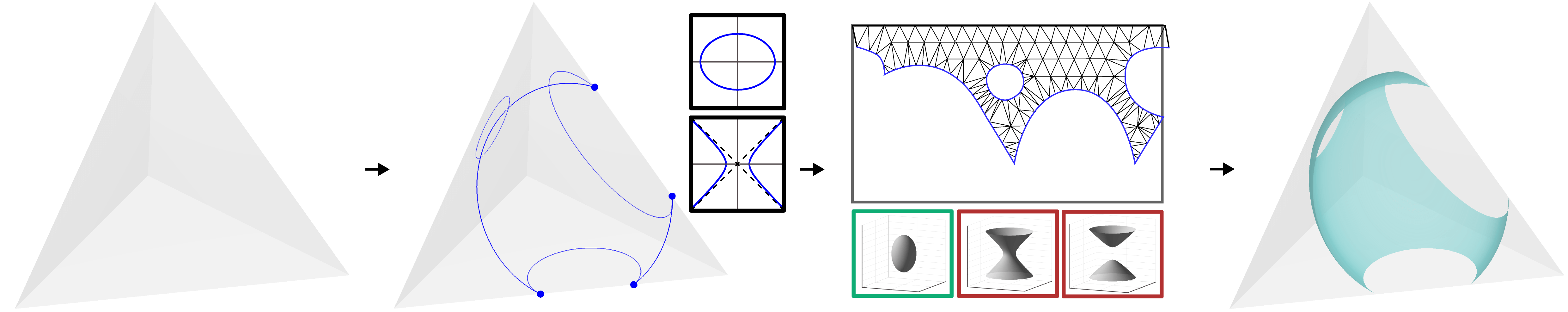}
				\put(55, -7)   {(a)}
				\put(30, 40)  {\LARGE{$f(x,y,z)$}}
				\put(180, -7)  {(b)}
				\put(325, -7)   {(c)}
				\put(455, -7)   {(d)}
			\end{overpic}
			\caption{%{\bf Quadratic surface extraction.}
				Given a quadratic function $f(x, y, z)$ defined on a tetrahedron (a), we find its zeroth levelset by first extracting the quadratic curves as an ellipse or a hyperbola on each face of the tetrahedron (b).
				These curves are then mapped to the parameter domain (c) for the quadratic surface (ellipsoid or hyperboloid) to bound a region, which we triangulate.
				Mapping the triangulated region back to the $XYZ$ space produces the quadratic surface (d).
			}
			\label{fig:method_quadratic}
		\end{figure*}
		\section{Extraction of Feature Curves and Surfaces}
		\label{sec:extraction}
		The input data to our visualization is a piecewise linear tensor field defined on a tetrahedral mesh. To extract the aforementioned feature curves and surfaces, we consider their complexity.
		Inside a tetrahedron, the tensor field is linear and can be locally expressed as $T(x, y, z) = T_0+xT_x+yT_y+zT_z$ where $T_0$, $T_x$, $T_y$, and $T_z$ are 3D asymmetric tensors.

		\subsection{Magnitude Surface}

The magnitude surface of a given magnitude $s>0$ is thus characterized by $||T||^2=s^2$. Note that when $T(x, y, z)$ is linear, $f(x, y, z)=||T(x, y, z)||^2-s^2$ is a quadratic polynomial of $x$, $y$, and $z$.
		Under structurally stable conditions, a quadratic surface is part of either an ellipsoid, a single-sheet hyperboloid, or a double-sheet hyperboloid.
		Note that each of these types of surfaces can be parameterized over two variables~\cite{hilbert:1999:geometry}. Therefore, we first express $f(x, y, z)$ as $f(x, y, z)= \textbf{x}^T K \textbf{x} = 0$ where $\textbf{x} = (x, y, z, 1)$ and  $K$ is a $4 \times 4$ symmetric matrix.
		Using the eigenvalues of $K$, we can decide whether the magnitude surface is an ellipsoid, a single-sheet hyperboloid, or a double-sheet hyperboloid~\cite{beyer:1991:crc}.
		Finally, we can extract the magnitude surface using proper parameterization at any given accuracy.
		Observing that the magnitude surface is piecewise quadratic and continuous across the faces of adjacent tetrahedra, we first extract the intersection of the magnitude surface with each face. Such an intersection is a quadratic curve inside each triangle face, which is part of either an ellipse or a hyperbola.
		We extract these quadratic curves using the method described in~\cite{Khan:20}.
		Next, we extract the magnitude surface inside each tetrahedron by collecting its intersection with the four faces of the tetrahedron.
		These intersection curves are then mapped to the parameter space for the magnitude surface (an ellipsoid or a hyperboloid), which form loops and bound a number of regions in the parameter space.
		We then sample each region at a given sampling rate to generate a set of points inside.
		Finally, we apply constrained Delaunay triangulation~\cite{chew:1989:constrained} with the boundary curves as the constraints to generate a triangulation of the regions in the parameter space.
		These regions, when mapped back to the $XYZ$-space, give rise to the magnitude surfaces in the tetrahedron. Fig.~\ref{fig:method_quadratic} illustrates this process. Finally, the magnitude surface from adjacent tetrahedra are stitched together from their shared quadratic curves in the common face.
		Note that the process of extracting the magnitude surface inside each tetrahedron is independent of the other tetrahedra.
		Thus, we enable parallel computation to speed up the process.
		\subsection{Isotropicity Surface}

The isotropicity surface is defined in Equation~\ref{eq:isotropicity}, which involves radicals in its formulation. To overcome this issue, we use an alternative formulation:
		\begin{equation}\label{eq:isotropicity_2}
			\trace(T)^2-3\eta(T)^2||T||^2=0
		\end{equation}
		Note that this formulation captures both the positive isotropicity surface and the negative isotropocity surface, and we refer to the collection of both surfaces as the {\em generalized isotropocity surface}.
		Such a surface is a quadratic surface in the domain, which we can extract using the same approach for extracting the magnitude surface mentioned above.
		The only issue is how to separate the positive and negative parts of the generalized isotropicity surface. This is achieved as follows.
		When extracting the intersection of the positive or negative isotropicity surface with a face of a tetrahedron, we use the sign of $\trace(T)$ to extract only the relevant intersection segments.
		We then use these segments as input to the remainder of our pipeline to extract the isotropicity surface inside each tetrahedron.
		This leads to the correct extraction of the surface, either positive only, or negative only.

		\subsection{Balanced Surface}
The balanced surface satisfies that $\minor(T)=0$.
		This is again a quadratic surface, which we extract using the same method as above.
		It consists of both the linear part and planar part, separated by the triple degenerate curve.
		We will provide the detail of extracting triple degenerate curves next, as part of our effort to extract neutral surfaces.

		\subsection{Neutral Surface and Triple Degenerate Curve}

The neutral surface satisfies $\det(T)=0$ and is a cubic surface.
		To extract such a surface, we employ the A-patches technique~\cite{luk:2009}, which allows the extraction of algebraic curves and surfaces.
		This is achieved by converting a degree-$n$ polynomial $f(x,y,z)$ into its Bernstein coefficients and testing the sign of coefficients on a tetrahedral grid to find the zeroth levelset.
		In addition, note that the triple degenerate curve is precisely the intersection of the neutral surface and the balanced surface. Consequently, we extract the triple degenerate curve as follows.
		We first extract the balanced surface using our quadratic surface extraction algorithm, which results in a triangular mesh. Next, we compute the curve $\det(T)=0$ on this mesh, which is the triple degenerate curve.
		To find this cubic curve on the triangular mesh, we employ the same A-patches method for a lower-dimension. That is, on a triangular mesh representing the balanced surface, we build a triangular Bernstein grid for each triangle, test the A-patch conditions~\cite{luk:2009}, and either extract the curve inside the triangle or subdivide the triangle into smaller triangles and repeat the process.
		%
		%More specifically, given a triangle in the mesh, we compute the Bernstein coefficients of $\det(T)=0$ for the triangle and place them on a triangular grid (Figure~\ref{fig:method_seg}). We then perform the same A-patch test for the triangular grid to determine whether there is an intersection of the curve with this triangle. If there is, we can either extract it directly when there is a separating layer in the grid or subdivide the mesh when there is not such a layer.
		%
		The ability to extract triple degenerate curves allows us to separate real neutral surfaces from complex neutral surfaces as well as separate linear balanced surfaces from planar balanced surfaces.
		
		\subsection{Degenerate surface and Mode surface}

 Other than the balanced surface and the neutral surface, all other mode surfaces are degree-six surfaces, including the degenerate surface.
		Such a surface can be extracted using the A-patches method.
		In addition, for such a surface, the linear part and the planar part are separated by precisely the triple degenerate curve.
		Thus, we can extract either the linear part, or the planar part, or both for any mode surface.
		\section{Performance}
		Our feature extraction algorithm is tested on a number of analytical and simulation data from solid mechanics and fluid dynamics. The number of tetrahedra in our data ranges from $500,000$ to $1,500,000$.
		Measurements were taken on a computer with an Intel(R) Xeon(R) E3-2124G CPU$@$ 3.40 GHz, 16GB of RAM, and an NVIDIA Quadro P620 GPU.
		The time to extract quadratic surfaces such as magnitude surfaces, isotropicity surfaces, and balanced surfaces range from $0.38$ second to $2.91$ seconds, depending on the number of tetrahedra in the data.
		It is more expensive to extract feature surfaces using the A-patches algorithm due to the recursive nature of the technique. The neutral surface is a degree-three surface. The time to extract this surface ranges from $0.69$ second to $5.90$ seconds. On the other hand, the degenerate surface and other mode surfaces are degree-six surfaces. The A-patches method requires $4.76$ seconds to $22.75$ seconds for our data.
		Note that the time reported above includes the time to compute the triple degenerate curve.

		\section{Applications}
		\label{sec:apps}
		In this section, we apply our novel analysis to a number of analytical and simulation datasets in solid mechanics and fluid dynamics.
		Additionally, we provide some physical interpretation of our visualization based on our tensor field analysis.
		\begin{figure}[!b]
			\centering%
			\begin{overpic}[width={0.95\columnwidth}]{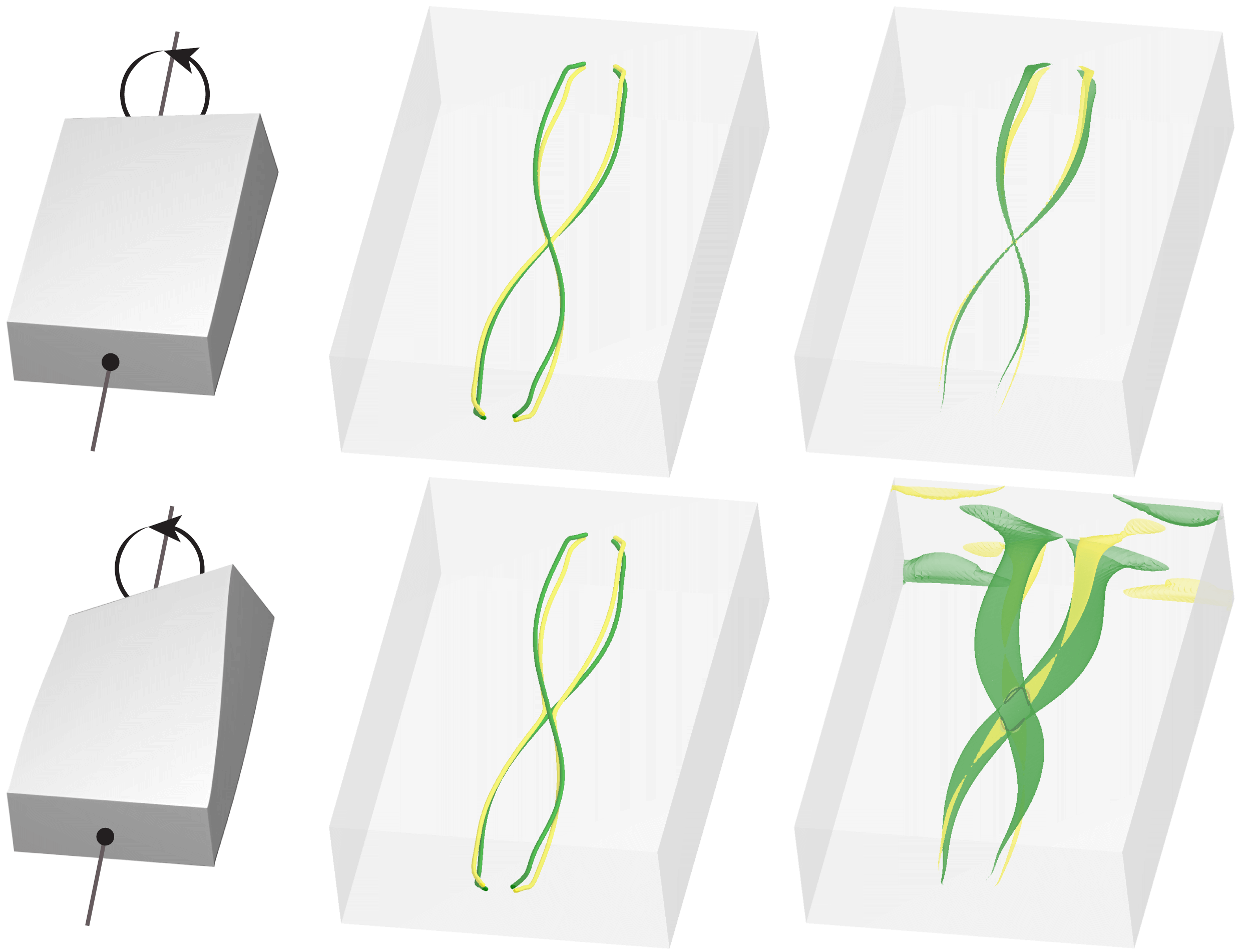}
				\put(-5, 165) {30\%}
				\put(-5, 70)  {100\%}
				\put(-5, -6) {(a) deformation}
				\put(70, -6)  {(b) Cauchy stress}
				\put(165, -6) {(c) PK1 stress}
			\end{overpic}
			\caption{%{\bf Cauchy and PK1 stress tensor fields.}
				We visualize the features at $30\%$ and $100\%$ of the loading of the twisting scenario.
				The images from the left to the right column are (a) the deformation, (b) the degenerate curves of the Cauchy stress tensor fields, and (c) the degenerate surfaces of the PK1 stress tensor fields.
			}
			\label{fig:app_Twist2}
		\end{figure}
		\begin{figure}[!b]
			\centering%
			\begin{overpic}[width={\columnwidth}]{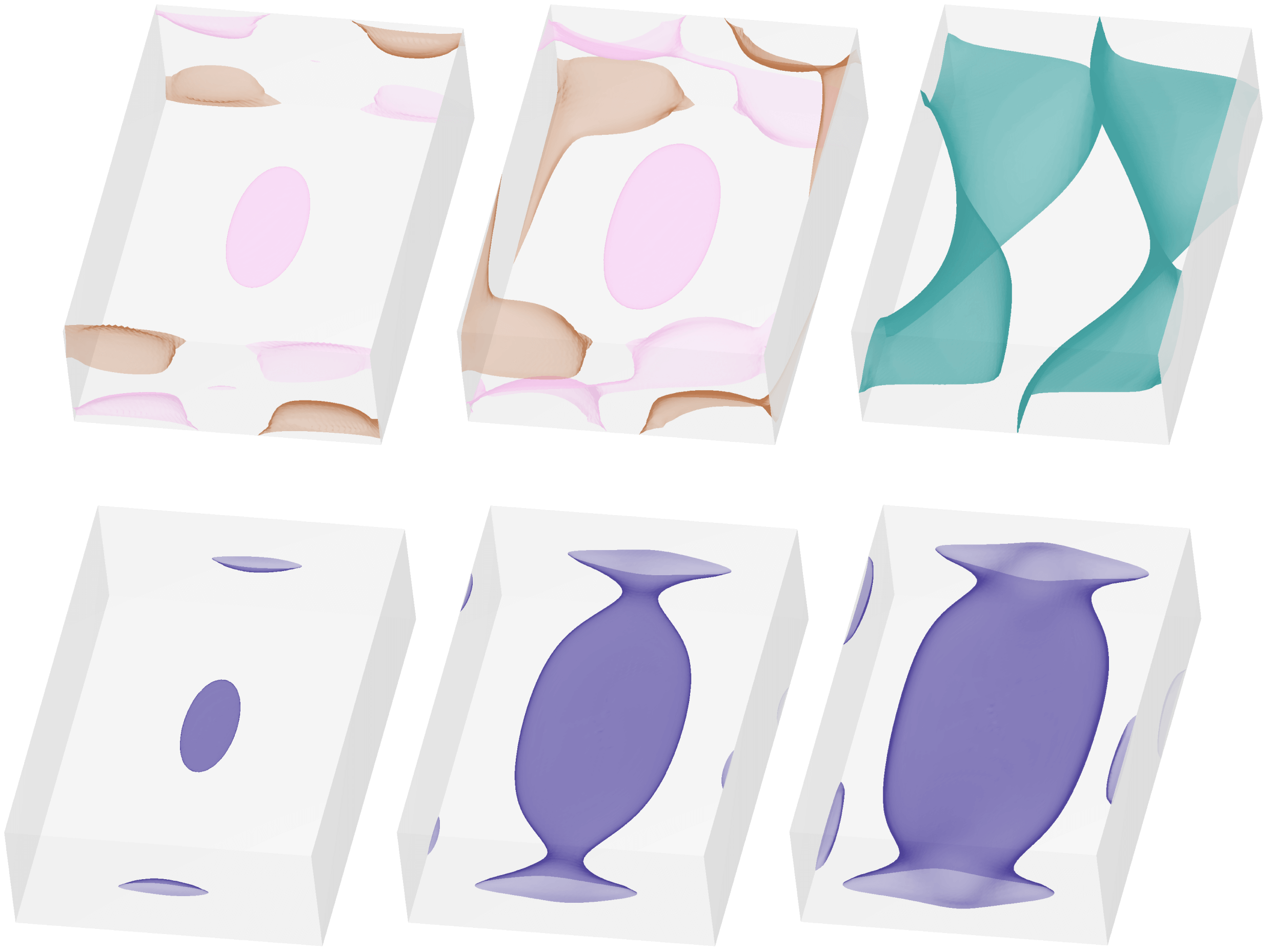}
				\put(15, 92)  {(a) $\eta(T)=\pm 0.7$}
				\put(90, 92) {(b) $\eta(T)=\pm 0.5$}
				\put(175,92) {(c) $\eta(T)=0$}
				\put(15, -6)  {(d) $\Vert T \Vert=50$}
				\put(90, -6) {(e) $\Vert T \Vert=100$}
				\put(175,-6) {(f) $\Vert T \Vert=150$}
			\end{overpic}
			\caption{%{\bf The non-topological feature surfaces of the twisting scenario.}
				We visualize three isotropicity surfaces (a-c) and three magnitude surfaces (d-f).
				The magnitude surfaces are colored in navy blue, while the positive and negative isotropicity surfaces are colored in brown and pink, respectively. The traceless surface (zero isotropicity) is colored in teal.
				%The symmetry in the twisting motion is also present in the isotropicity surfaces and magnitude surfaces.
			}
			\label{fig:app_Twist}
		\end{figure}
		\subsection{Solid Mechanics}
		Twisted bundles of steel cables can be found in many places in real life, from those embedded in truck tires to bring additional support, to those used for suspension structures such as cable cars, elevators, and cranes (machines). The cables can fail due to the wear and tear from the cables untwisting under stress (heavy weight lifting).
		Such failures can in turn lead to property damages and loss of lives.
		To understand the potential weakness in the steel cables under stress due to twisting, we consider the first Piola-Kirchhoff (PK1) stress tensor~\cite{kelly:2020:solid} used to study metal plasticity.
		Unlike the perhaps better-known Cauchy stress which is symmetric, the PK1 stress tensor is asymmetric as it is the product of the Cauchy stress and the deformation gradient tensor.

		{\bf The twisting scenario}
		is simulated with SIMULIA~\cite{ABAQUS:614}.
		Figure~\ref{fig:app_Twist2} shows two twisting stages of a block whose front face is fixed and the back face is twisted: (top) $30\%$ of the full twist, and (bottom) fully twisted ($18^\circ$).
		We observe that the degenerate curves in the Cauchy stress tensor fields (Figure~\ref{fig:app_Twist2} (b): colored curves) and the degenerate surfaces in the PK1 stress tensor fields (Figure~\ref{fig:app_Twist2} (c): colored surfaces) both have a twisting structure.
		However, the Cauchy stress tensor fields (Figure~\ref{fig:app_Twist2} (b)) do not have a complex domain and the set of degenerate tensors in the field form curves. This is visible from the visualization shown in the middle column, where linear degenerate curves are colored in green and planar degenerate curves are colored in yellow.
		Note that despite the significantly different twists at different stages, the Cauchy stress leads to the nearly identical set of degenerate curves (Figure~\ref{fig:app_Twist2} (b): top and bottom).
		In contrast, the visualization of the PK1 tensor fields (Figure~\ref{fig:app_Twist2} (c)) shows a clear difference in the degenerate surfaces for the two stages. While at $30\%$ (Figure~\ref{fig:app_Twist2} (c): top), the degenerate surface in the PK1 tensor is similar to the degenerate curves in the Cauchy stress (Figure~\ref{fig:app_Twist2} (b): top), at $100\%$, a pronounced difference is shown (Figure~\ref{fig:app_Twist2} (b-c): bottom).
		This highlights the potential benefits of visualizing the asymmetric PK1 stress over the symmetric Cauchy stress for twisting motions.
		Moreover, in the PK1 stress tensor fields, the linear and planar degenerate surfaces point out the boundary conditions of the fixed side and the twisting side of the block.
		Since the fixed side has less deformation, the region of the complex domain is smaller.
		Furthermore, while the loading is increasing, the complex region is growing, which indicates that the rotation-dominant domain is getting larger.

		Figure~\ref{fig:app_Twist} visualizes the isotropicity surfaces and the magnitude surfaces which have a skew-symmetric structure as well.
		In addition, the isotropicity surfaces illustrate the material is compressed at the center and isotropically stretched on the boundary. Insights such as these are dependent on the ability to perform tensor field analysis.

		Note that both the PK1 stress and the Cauchy stress can provide important insights into the underlying mechanics as such insights are complementary. As tensor fields, certain tools are available for both symmetric and asymmetric tensor fields, such as eigenvalue analysis. On the other hand, the interpretation of such analysis depends on many factors such as the type of the tensors and the physical quantities that they represent. Consequently, visualizing both types of tensor fields and understanding the connection between their structures, i.e. multi-field visualization, can provide a more holistic view of the underlying physics than using only one of them.

		\subsection{Fluid Dynamics}
		The velocity gradient tensor field of a flow plays an important role in understanding fluid dynamics, and asymmetric tensor field analysis of such a field can lead to complementary insight to existing vector field visualization methods~\cite{Zhang:09}.
		In this paper, we perform analysis and visualization of 3D velocity gradient tensor fields directly instead of their 2D projections onto some lower-dimensional surface or probe plane. We will discuss our data sets: (1) the Lorenz attractor, (2) the Rayleigh-B{\'e}rnard flow, (3) the Arnold–Beltrami–Childress flow, and (4) an open-channel flow (Appendix).
		{\bf Lorenz attractor} is a set of chaotic solutions to the Lorenz system~\cite{lorenz:1963:deterministic} with system parameters $\sigma$, $\rho$, and $\beta$. Figure~\ref{fig:teaser} (a) shows the butterfly-shaped attractor (the grey winding curve) in the system when $\sigma =10$, $\rho =28$, and $\beta =8/3$~\cite{lorenz:1963:deterministic}.
		We extract and visualize the feature curve and surfaces in the gradient tensor, such as} (b) linear degenerate surfaces (green) and planar degenerate surfaces (yellow), (c) real neutral surfaces (orange) and complex neutral surfaces (red), and (d) linear balanced surfaces (blue) and planar balanced surfaces (magenta).
	Note that all of these surfaces intersect exactly at the triple degenerate curves (black).
	Moreover, topological feature surfaces separate the two critical points in the attractor, and these surfaces exhibit a two-way rotational symmetry.
	
	%
	
	%%%%%%%%%%%%%%%%%%%%%%%%%%%%%%%%%%%%%%%%%%%%%%%%%%%%%%%%%%
	%
	\begin{figure*}[!t]
		\centering%
		\begin{overpic}[width={\textwidth}]{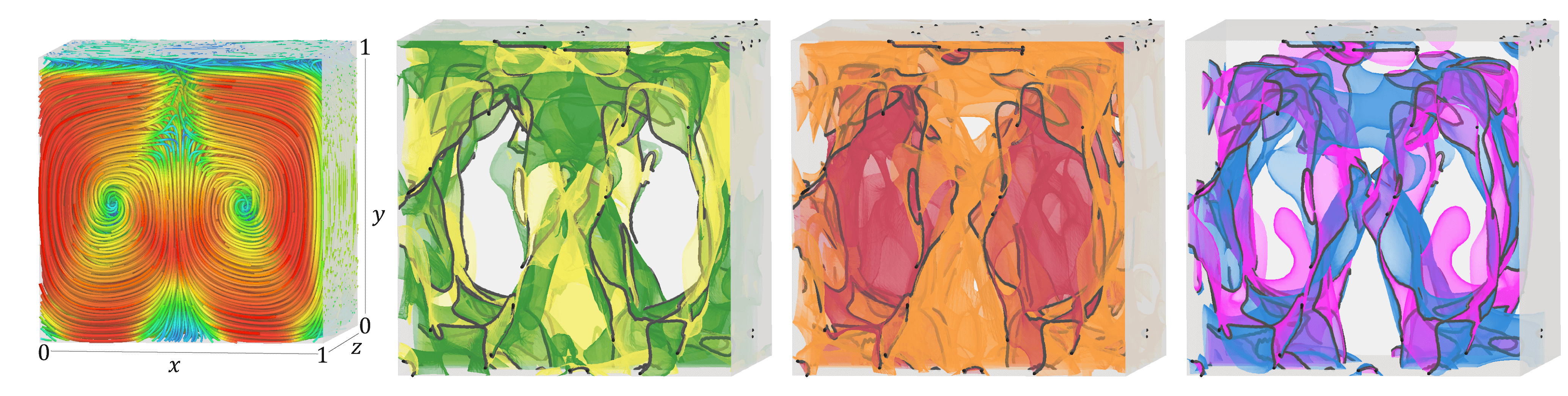}
			\put(25, -7)    { (a) vector field}
			\put(145, -7)  { (b) degenerate surfaces}
			\put(275, -7)  { (c) balanced surfaces}
			\put(410, -7)  { (d) neutral surfaces}
		\end{overpic}
		\caption{
			This figure visualizes the Rayleigh-B{\'e}rnard flow with (a) the vector field, (b) the linear and planar degenerate surfaces (green and yellow), (c) the real and complex neutral surfaces (orange and red) neutral surfaces, and (d) linear and planar balanced surfaces (blue and magenta).
			Notice that the triple degenerate curve (black curves) can be considered as the spine to which the feature surfaces are attached.
		}
		\label{fig:app_Bernard}
	\end{figure*}
	\begin{figure}[!t]
		\centering%
		\begin{overpic}[width={\columnwidth}]{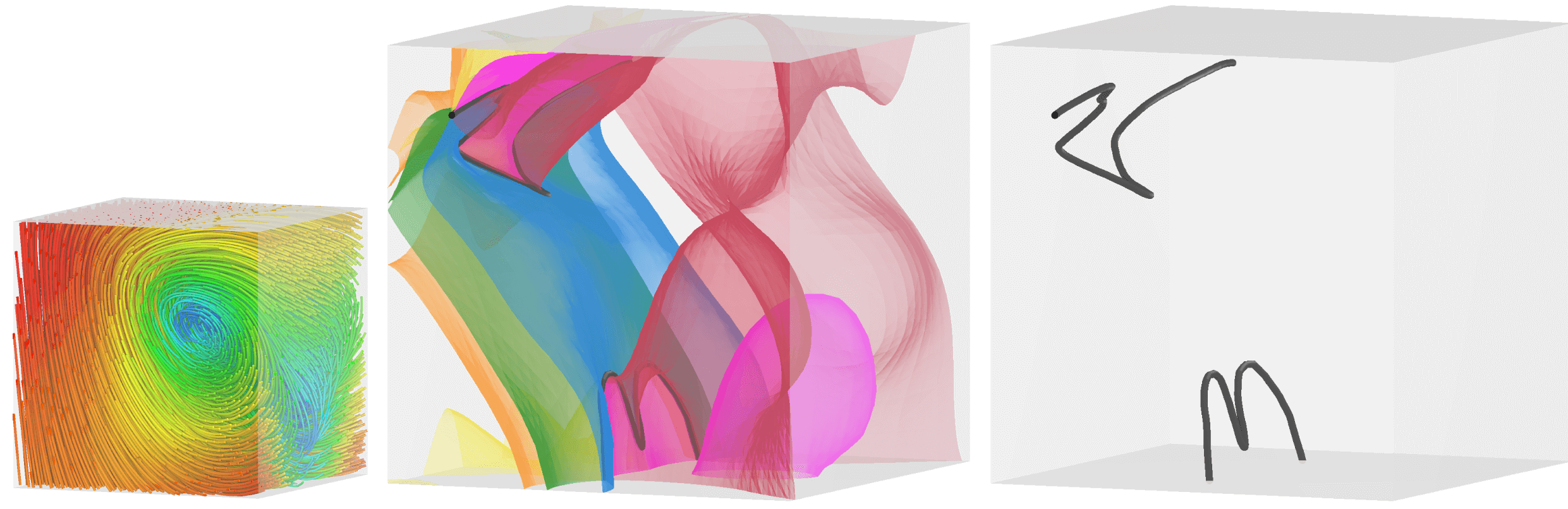}
			\put(0,  -8) {(a) vector field}
			\put(75, -8) {(b) mode surfaces}
			\put(150, -8) {(c) triple degenerate curves}
		\end{overpic}
		\caption{%
			We magnify the center portion of the B{\'e}rnard cell and show (a) the streamlines of the vector field, (b) the mode surfaces and (c) the triple degenerate curves.
			%
			%In  (b), we also show the vortex core lines (golden curves) which are in the rotation-dominant domain.
		}
		\label{fig:app_Bernard2}
	\end{figure}
	
	{\bf The Rayleigh-B{\'e}rnard flow} is thermal convection in a thin horizontal layer of fluid heated from below by maintaining the constant temperature difference between the upper and lower boundaries.
	The flow is characterized by the formation of two convection cells as shown in Figure~\ref{fig:app_Bernard} (a).
	Identifying regions of stretching and compression as well as the rotation-dominant region is useful yet challenging.
	With our feature surfaces, such spaces can be better perceived. Notice that the triple degenerate curve (black) can be considered as a spine to which other feature surfaces are attached.

	In Figure~\ref{fig:app_Bernard2} (a), we zoom in on the center portion of the B{\'e}rnard cell ($x \in [0.5, 0.8]$; $y \in [0.35, 0.65]$; $z \in [0.35, 0.65]$) and visualize the neutral surfaces, the degenerate surfaces, the balanced surfaces, and the triple degenerate curves (Figure~\ref{fig:app_Bernard2} (b)).
	Note that the left face of the cube corresponds to the center face that separates the pair of convection cells and the lower-left corner contains the converging upwelling flow.
	There, we observe a quick transition of the relatively flat and parallel mode surfaces: the planar degenerate surface (leftmost, yellow), the real neutral surface (orange), the linear degenerate surface (green), and the linear balanced surface (rightmost, blue).
	Near the bottom, underneath the real neutral surface, the converging flow is dominated by shearing with compression (planar), and then becomes stretching-dominant (linear) by crossing the real neutral surface.
	Next, the strength of the rotation gradually increases until the shear balances the rotation at the linear balanced surface (blue).
	Finally, we enter the rotation-dominant convection cell domain on the right-hand side of the linear balanced surface (blue).
	%
%	We also observe that there is another flow-pattern transition under the triple degenerate curve: the planar degenerate (yellow), the planar balanced (magenta) and then leading to the rotation-dominant domain.
	%

	Furthermore, in the upper part of upwelling convection, the flow characteristic transitions exhibit more volumetric appearance: from the linear degenerate surface (leftmost, green), real neutral surface (orange), planar degenerate surface (yellow), planar balanced surface (magenta), to the complex neutral surface (rightmost, red) in the rotation cell.
	Lastly, we notice both the triple degenerate curves have an ``M'' shape, with the triple degenerate curve near the bottom of the cube being narrower (Figure~\ref{fig:app_Bernard2} (c)).
	%
	%They are symmetric with respect to the y-axis, which is a reflection due to the symmetry in the flow field.
	%
	We conjecture that the converging flow pattern pushes the triple degenerate curves towards the center of the domain, thus the ``M'' shapes.
	The above observations of the flow characteristics and behaviors can be difficult to detect and interpret correctly with the 2D flow and tensor field visualization in probe planes.
	Such comprehensive analysis is attainable with the use of 3D visualization of the velocity gradient tensors.

		\begin{figure*}[!t]
			\centering%
			\begin{overpic}[width={\textwidth}]{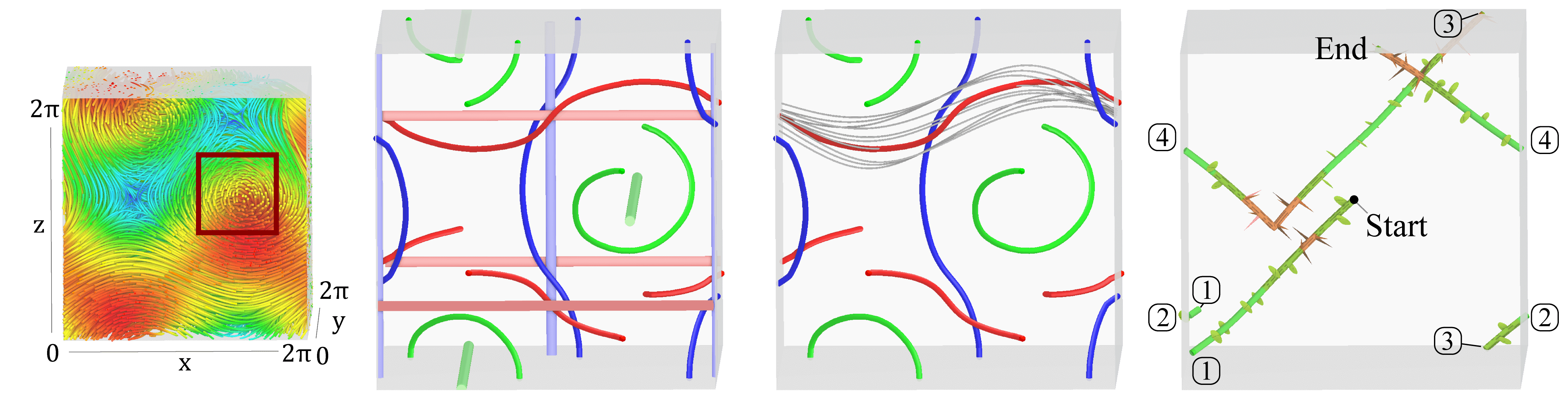}
				\put(30, -7)  {(a) Vector field}
				\put(170, -7)  { (b) }
				\put(310, -7)  { (c) }
				\put(435, -7)  { (d) }
			\end{overpic}
			\caption{On the boundary face of the ABC flow, streamlines exhibit {\em ordered} behavior (a: inside the red frame) and {\em chaotic} behaviors (a: outside the red frame). The non-chaotic streamlines occur around the cylindrical regions (b: transparent cylinders and half-cylinders) that are referred to as the principal vortices. Notice the one-to-one correspondence between the principal vortices and the triple degenerate curves (b: colored curves). In addition, the chaotic streamlines (c: grey curves) appear to travel along the corresponding triple degenerate curve. The dominant hyperstreamline starting from $(\pi, \pi, \pi)$ (d) contains multiple real segments (with thorns) and complex segments (with leaves) as it travels in the periodicity box. Note that when intersecting a face, the hyperstreamline continues from the same location on the opposite face, i.e. points with the same labels. Successive intersection points are labeled with $1$-$4$. Notice that the hypersteamline is mostly a straight line except when it crosses the real neutral surface (not shown). It also consists of mostly complex segments.
			}
			\label{fig:app_ABC}
		\end{figure*}

	%%%%%%%%%%%%%%%%%%%%%%%%%%%%%%%%%%%%%%%%%%%%%%%%%%%%%%%%%%
	{\bf Arnold–Beltrami–Childress flow} (ABC flow) is a 3D incompressible vector field that is a steady-state solution to Euler's equations~\cite{dombre:1986:chaotic}. The ABC flow is periodic in each of the $X$, $Y$, and $Z$ directions with a period of $2\pi$ and is usually studied in its {\em periodicity box}: $[0, 2\pi) \times [0, 2\pi) \times [0, 2\pi)$ (Figure~\ref{fig:app_ABC} (a): the cube). One of the main characteristics of the ABC flow is the existence of {\em chaotic streamlines}, which,  due to the periodicity in the flow, can intersect a face of the periodicity box infinitely many times so that the set of the intersection points fills a region in the face~\cite{dombre:1986:chaotic}. When $A=1$, $B=\sqrt{2/3}$, and $C=\sqrt{1/3}$, chaotic streamlines occur outside the so-called {\em principal vortices}, each of which is a tubular region along one of the $X$, $Y$, and $Z$ axes (Figure~\ref{fig:app_ABC} (b): colored cylinders and half-cylinders due to periodicity). There are a total of six principal vortices, two along each axis. Inside a principal vortex, the streamlines' orientations are predominantly along the direction of the tube. Each such streamline intersects a face of the cube at a set of points that are on a curve (instead of a region). Such streamlines are not chaotic. On the faces of the periodicity box, the intersection points with chaotic streamlines are outside the principal vortices. While Dombre and Frisch~\cite{dombre:1986:chaotic} illustrate the principal vortices as cylinders, they point out that these regions are helical, which, when traveling from one face to the opposite face of the cube, finish a turn of $2\pi$. We observe that there is a one-to-one correspondence between the set of triple degenerate curves (Figure~\ref{fig:app_ABC} (b): colored curves) and the set of principal vortices. Note that some of the triple degenerate curves are divided into three segments by the faces of the periodicity box. We notice that each triple degenerate curve also has a helical shape and finishes a turn of $2\pi$ after traveling from one face to the opposite face. Moreover, each triple degenerate curve is a loop under the periodic condition. In addition, the streamlines in a principal vortex (Figure~\ref{fig:app_ABC} (c): grey curves) appear to be around the triple degenerate curves. The correlation between the triple degenerate curves and the principal vortices in terms of their numbers, locations, and shapes suggests that additional insights may be gained on the ABC flow by inspecting the topological structures in its gradient tensor field.

	\begin{figure}[!t]
		\centering%
		\begin{overpic}[width={\columnwidth}]{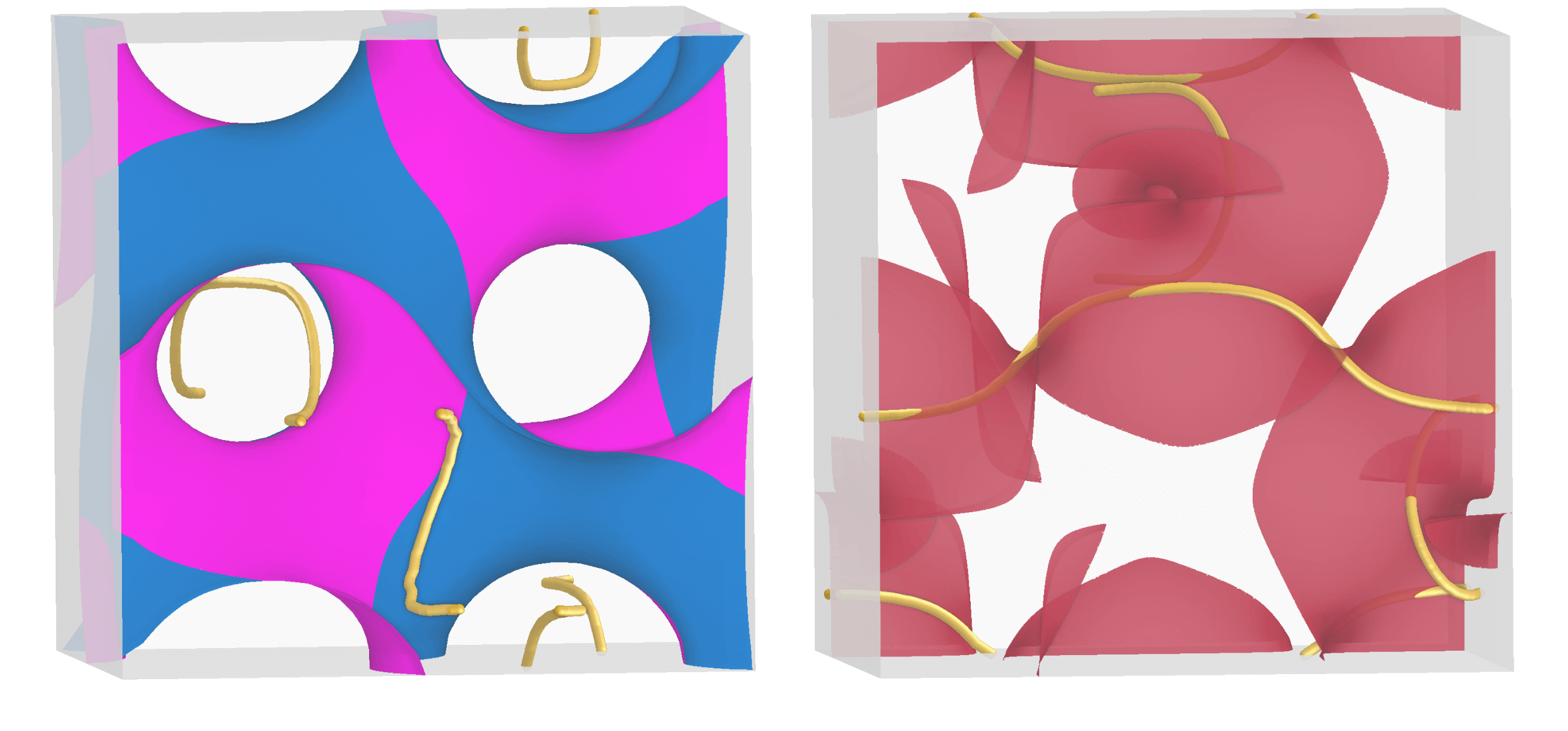}
			\put(25, 0)   { (a) Balanced surfaces}
			\put(50, -7)   { \small (back view)}
			\put(135, 0) { (b) Complex neutral surfaces}
			\put(175, -7)   { \small (side view)}
		\end{overpic}
		\caption{
			We compare the vortex core lines (golden curves) of the ABC flow to (a) the linear and planar balanced surfaces (blue and magenta), and (b) the complex neutral surfaces (red).
			Notice that the vortex core lines appear to be in the inner complex domain (a) and separated into the linear and planar segments by the complex neutral surfaces (b).
		}
		\label{fig:app_ABC2}
	\end{figure}

	Besides the periodicity in the flow, there is an additional eight-fold symmetry within the periodicity box~\cite{dombre:1986:chaotic} that leads to the {\em fundamental box}: $[0, \pi) \times [0, \pi) \times [0, \pi)$ which is one-eighth of the periodicity box. Given this, we compute the dominant hyperstreamline from $p=(\pi, \pi, \pi)$ in one direction.
%Note that the vector value at $p$ is $(-C, -A, -B)$ where $A$, $B$, and $C$ are the parameter value. In addition, $p$ is on the balanced surface.
Figure~\ref{fig:app_ABC} (d) shows the augmented hyperstreamline through $p$, which, when intersecting a face of the periodicity box, continues from the same location on the opposite face. Successive intersection points are labeled with $1$-$4$. Notice that the hyperstreamline is mostly straight, except where the dominant eigenvectors are discontinuous (crossing the real neutral surfaces). The variety of tensor field behavior along the hyperstreamline reflects the rich structure in the ABC flow. This highlights the benefit of our tree-based augmented hyperstreamline visualization, which can be used as a probing tool for the field with only one user-specified seed point. The eigenvector information in the field along the hyperstreamline is captured by the thorns and leaves, giving the user a more holistic view of the field than showing only the stem.
	
	Vortex core lines are a popular visualization for understanding fluid flows~\cite{roth:1998:higher}. We compare vortex core lines (extracted using VTK~\cite{schroeder:2006:VTK} to our tensor-based feature surfaces. As shown in Figure~\ref{fig:app_ABC2} (a) (colored curves), there are four vortex core lines given the periodicity condition. Notice that they do not intersect the balanced surfaces (a) but intersect the complex neutral surfaces (b). This suggests that the vortex core lines are inside not only the complex domain but also the inner complex domain. Such an observation signifies the importance of the balanced surfaces in understanding fluid flows. In addition, if this observation can be justified theoretically, it may be used in the future to evaluate the effectiveness of vortex core line extraction methods. Moreover, the complex neutral surface divides vortex core lines into linear segments and planar segments. To our knowledge, a vortex core line is usually extracted and studied as a whole. Understanding the transition from linear parts to planar parts and vice versa has the potential of bringing additional insight to the understanding of the underlying fluid dynamics.

	\section{Conclusion and Future Work}
	\label{sec:conclusion}

	In this paper, we explore the topology of 3D asymmetric tensor fields and introduce an eigenvalue space based on the tensor mode that facilitates our analysis.
	At the core of our analysis is the definition of tensor mode, which gives rise to a number of feature curves and surfaces with topological significance. In addition, we show that triple degenerate tensors are stable and form curves.
	Additionally, we introduce the notion of balanced surface, which divides the complex domain into the inner part (rotation-dominant) and the outer part (shear-dominant). Such a feature is not present for 2D asymmetric tensor fields.
	%
	%	We also provide the topological significance of our feature surfaces by studying the tensor behavior when crossing these surfaces.
	%
	Observing that a number of the feature surfaces are quadratic, we provide an algorithm to extract them effectively and quickly. Note that our algorithm can also be used to extract quadratic feature surfaces in symmetric tensor fields such as the magnitude surfaces and the isotropic index surfaces~\cite{Palacios:16}.
	To enable a holistic view of the eigenvectors and dual-eigenvectors, we visualize a hyperstreamline following one eigenvector field as a tree stem with attached thorns and leaves to show the other eigenvectors or dual-eigenvectors.
	This allows us to inspect the change in the tensor field behavior across important feature surfaces.
	Finally, we have applied our analysis and visualization to a number of analytical and simulation data and provided some physical observations.
	In the future, we wish to investigate more robust extraction of feature surfaces of the tensor fields than the A-patches method, which neither guarantees to find all the surfaces nor provides a seamless surface extraction.
	For example, there has been work on the seamless extraction of mode surfaces for 3D symmetric tensor fields~\cite{Qu:21}, which is based on a reparameterization of the space of mode surfaces in a linear tensor field.
	We plan to investigate a potential adaptation of this approach for asymmetric tensor fields.
	Deeper understanding of the relationship between features in a 3D asymmetric tensor field and those of its symmetric part is another direction that we plan to explore.
	The two types of tensors share many characteristics such as the concepts of eigenvalues and eigenvectors as well as tensor invariants such as the magnitude and trace.
	Moreover, the symmetric part of an asymmetric tensor is symmetric, and understanding how the topological features in an asymmetric tensor field such as neutral surfaces, degenerate surfaces, and balanced surfaces relate to the features in the symmetric part has the potential of creating a unified framework for 3D tensor fields, whether symmetric or asymmetric.
	Such deeper understanding can potentially lead to more insight into the data by examining the features in both the asymmetric tensor field and its symmetric part.
	Finally, developing a multi-scale representation of 3D asymmetric tensor field topology is an important research area that has received relatively little attention from the Visualization community.
	We plan to investigate this area in our future work.
	%

	% if have a single appendix:
	%\appendix[Proof of the Zonklar Equations]
	% or
	%\appendix  % for no appendix heading
	% do not use \section anymore after \appendix, only \section*
	% is possibly needed
	
	% use appendices with more than one appendix
	% then use \section to start each appendix
	% you must declare a \section before using any
	% \subsection or using \label (\appendices by itself
	% starts a section numbered zero.)
	%
	%
	%\appendices
	%
	%\section{Proofs}
	%\label{sec:proofs}}
	%The appendix will be eventually completely absorbed in the main text. They are here for reference purposes only.
	%
	%
	\acknowledgments{The authors wish to thank our anonymous reviewers for their constructive feedback. We appreciate the help from Avery Stauber and Yichuan Yin during video production. This work was supported in part by the NSF award (\#1619383).
	}

	\bibliographystyle{abbrv-doi}

	\bibliography{3DAsymmetric}
	
\end{document}